\documentclass[preprint,11pt]{JHEP3}

\JHEPspecialurl{http://jhep.sissa.it/JOURNAL/JHEP3.tar.gz}

\usepackage{epsfig,multicol,amsmath}

\usepackage{epstopdf}
\usepackage{bm}  
\usepackage{amsmath}     
\usepackage{epsfig,epsf}
\usepackage{showkeys}
\usepackage{graphicx}
\usepackage{rotating}
\usepackage{cite}

\def\beq{\begin{equation}}
\def\eeq{\end{equation}}
\def\bea{\begin{eqnarray}}
\def\eea{\end{eqnarray}}

\def\eq#1{{Eq.~(\ref{#1})}}
\def\fig#1{{Fig.~\ref{#1}}}

\newcommand{\Lb}{\left(}
\newcommand{\Rb}{\right)}
\parskip=\itemsep               

\newcommand{\nn}{\nonumber}

\newcommand{\h}{\frac{1}{2}}

\newcommand{\A}{{\cal A}}

\newcommand{\p}{I\!\!P}

\def\pom{{I\!\!P}}

\title{Survival  probability of large rapidity gaps in the QCD and N=4 SYM motivated model.}
\author{\Large  E. Gotsman$^{a}$\thanks{Email:
gotsman@post.tau.ac.il.}\,, E. Levin$^{a,b}$\thanks{Email:
leving@post.tau.ac.il}\,\,and\,\,U. Maor$^{a}$\thanks{Email: maor@post.tau.ac.il.}\, 
\\
a)\,  \,Department of Particle Physics, School of Physics and Astronomy,
Raymond and Beverly Sackler
 Faculty
of Exact Science,  Tel Aviv University, Tel Aviv, 69978, Israel\\
b)\,\,Departamento de F\'\i sica, Universidad T\'ecnica
Federico Santa Mar\'\i a, Avda. Espa\~na 1680,
Casilla 110-V,  Valparaiso, Chile 
\\}

\abstract{ 
	In this paper we 
present a self consistent theoretical approach for the  calculation of the 
  Survival Probability for central dijet production . These calculations 
 are performed in a model of
 high energy soft
 interactions based on two ingredients:(i) the results of N=4 SYM, which at the moment
  is the only theory 
 that is able to deal with a large coupling constant; and (ii) the required  matching with high energy 
QCD.
 Assuming, in accordance with these prerequisites, that soft Pomeron intercept is  rather large and  the 
slope
 of the Pomeron trajectory is equal to zero, we derive  analytical formulae that sum both enhanced
 and semi-enhanced diagrams for elastic and diffractive amplitudes. Using parameters obtained from a fit 
to the available
 experimental data, we calculate the Survival Probability for central dijet production at energies  
accessible at the LHC. The results presented here which   include the contribution of semi-enhanced 
and net diagrams,  are considerably larger than our previous 
estimates.

}

\dedicated{PACS: 13.85.-t, 13.85.Hd, 11.55.-m, 11.55.Bq}
\preprint{TAUP - 2011/2\\
{\tt 1101.5816 [hep-ph]}\\
\today}
\begin{document}

\section{Introduction}

  The original evaluation of the survival probability for a large rapidity gap (SPLRG) was made by 
Bjorken \cite{BJ}(see also Ref.\cite{DOK}), in a one channel formalism, where the elastic amplitude was assumed to have a
Gaussian shape in impact parameter space. This enabled him to obtain a closed analytic expression
 for $<|S^2|>$, the SPLRG.

We \cite{GLM1}  extended Bjorken's treatment to five different models, all of which reproduce the 
measured total cross sections at energies upto and including W = 1.8 TeV. However, the values obtain from 
these models for $<|S^2|>$  differed greatly, e.g. for the Tevatron energy they predicted values
between 9.6 and 32.6\%. 
Indicating that using parameters determined only from the elastic amplitude, was 
insufficient to determine $<|S^2|>$ uniquely.

We then broadened our approach to a two channel Good-Walker description, which included elastic as well 
as diffractive channels \cite{GLM2}. This was called for, as the experimental data from the Tevatron
\cite{Tevatron} indicated that the diffractive channels were important and should not be neglected, 
$\frac{\sigma_{sd}+ \sigma_{dd}}{\sigma_{elast}} \; \approx \;0.85$. Including the diffractive channels
necessitated including also the triple Pomeron coupling, which plays a crucial role in determining 
$<|S^2|>$.

The difficulty inherent in including the diffractive channels (i.e. 3$\pom$ coupling), is that one has 
to sum over all orders (loops) of these interactions, and there is no rigorous method for evaluating the 
so called enhanced and semi-enhanced (Pomeron) diagrams.

To make the problem tractable, we have in two previous papers (see Refs.\cite{GLMM,GLMLA}) 
 constructed a model of soft interactions at high 
energies based 
on the postulates that stem from N=4 SYM and QCD. The model includes the following  features:

\begin{enumerate}
\item \quad It is  built using Pomeron
 and Reggeons as the main ingredients as  follows from N=4 SYM( see Refs.
\cite{BST,HIM,COCO,BEP,LMKS});
\item \quad The intercept of the Pomeron should be rather large.
 In N=4 SYM\cite{BST,BFKL4} we 
expect $\Delta_{\p} = \alpha_{\p}(0) - 1 =
1 - 2/\sqrt{\lambda} \approx 0.11 \sim 0.33$, based on estimates of 
$\lambda \,=\,5 \sim 9$ derived from the early LHC measurements
 from the cross section of  multiparticle production
as well as from DIS at HERA \cite{LEPO}.

\item As follows from N=4 SYM, 
  the slope of the Pomeron \quad $\alpha'_{\p}(0) \,=\,0$, this is compatible with our fits to the cross 
section data in the ISR-Tevatron energy range \cite{GLMM,GLMLA}.

\item \quad  A large contribution to the cross section  comes from  processes
 of  diffraction dissociation, as in N=4 SYM at large $\lambda$
 only these processes 
 contribute to the scattering amplitude.
 In other words,  in this model
  the Good-Walker mechanism \cite{GW} is the main source
 of  diffractive production;
\item \quad  The Pomeron self-interaction should be small
 (of the order of $2/\sqrt{\lambda}$ in N=4 SYM), and much smaller than
 the vertex of interaction of the Pomeron with a hadron, which is
 of the order of $\lambda$;

\item \quad The last condition is not a prerequist of N=4 SYM, but follows from the requirement of the 
natural 
matching with perturbative QCD, where the only  vertex that  
contributes is the triple Pomeron vertex\cite{BART}.
 This ingredient differentiates our model from other models on the market (see Refs.\cite{KMRS,OS});
\end{enumerate}

 In this paper we   apply the  
 theoretical approach that was developed
to  calculate  the value for 
  SPLRG processes  at the LHC.
This problem has been a subject of much discussion \cite{GLMM,KMRS} since our  estimates first given in
 Ref.\cite{GLMM}, suggested a  small value of the survival probability for central Higgs boson 
production,
 in disagreement with the estimates of the  Durham group  \cite{KMRS}. 
In Ref. \cite{GLMM} we only included the enhanced Pomeron diagrams. 
 In this paper we present our estimates for the full set of the diagrams. We  discuss the
  important topic : the derivation of  formulae for 
 SPLRG processes originating  from three sources: the Good-Walker mechanism, semi-enhanced diagrams and
 enhanced diagrams for the Pomeron interaction.   We discuss this derivation in the first three
 sections, while the fourth section  is devoted to  numerical estimates.

\section{The main formulae: Good-Walker mechanism}
The calculation of the survival probability for the hard processes is a choice example that shows
 that the cross section of the hard processes cannot be calculated, except for some special
 cases (total DIS cross section, inclusive cross sections), without a substantial knowledge
 of the `soft' physics.  The general formula for  the Good-Walker mechanism (sum of the non-enhanced
 Pomeron diagrams) and for the sum of enhanced diagrams  has the form

\beq \label{SP}
\langle\mid S^2 \mid \rangle = \frac{\int \,\,d^2\,b\,\left\{\sum_{i,k} \,<p|i>^2 <p|k>^2\,e^{-\frac{1}{2} \Omega_{i,k}\Lb s,b\Rb}\,\,S^H_{i,k}(b)\right\}^2 }{\int\,d^2\,b\,
\left\{\sum_{i,k} <p|i>^2 <p|k>^2\,\int\,d^2 b'  A^i_H(s,b)\,A^k_H(s,\vec{b} - \vec{b}')\right\}^2},
\eeq
where,
$<p|i>$ is  equal to $\langle \Psi_{proton}\mid \Psi_i \rangle$ and , 
therefore, $<p|1>
= \alpha$ and $ <p|2> = \beta$, $\alpha^2 \; +\beta^2\;$ = 1. Our notation as well as the 
derivation of  
\eq{SP}
  can be found in Ref. \cite{2CH}.

\begin{figure}
\centerline{\epsfig{file=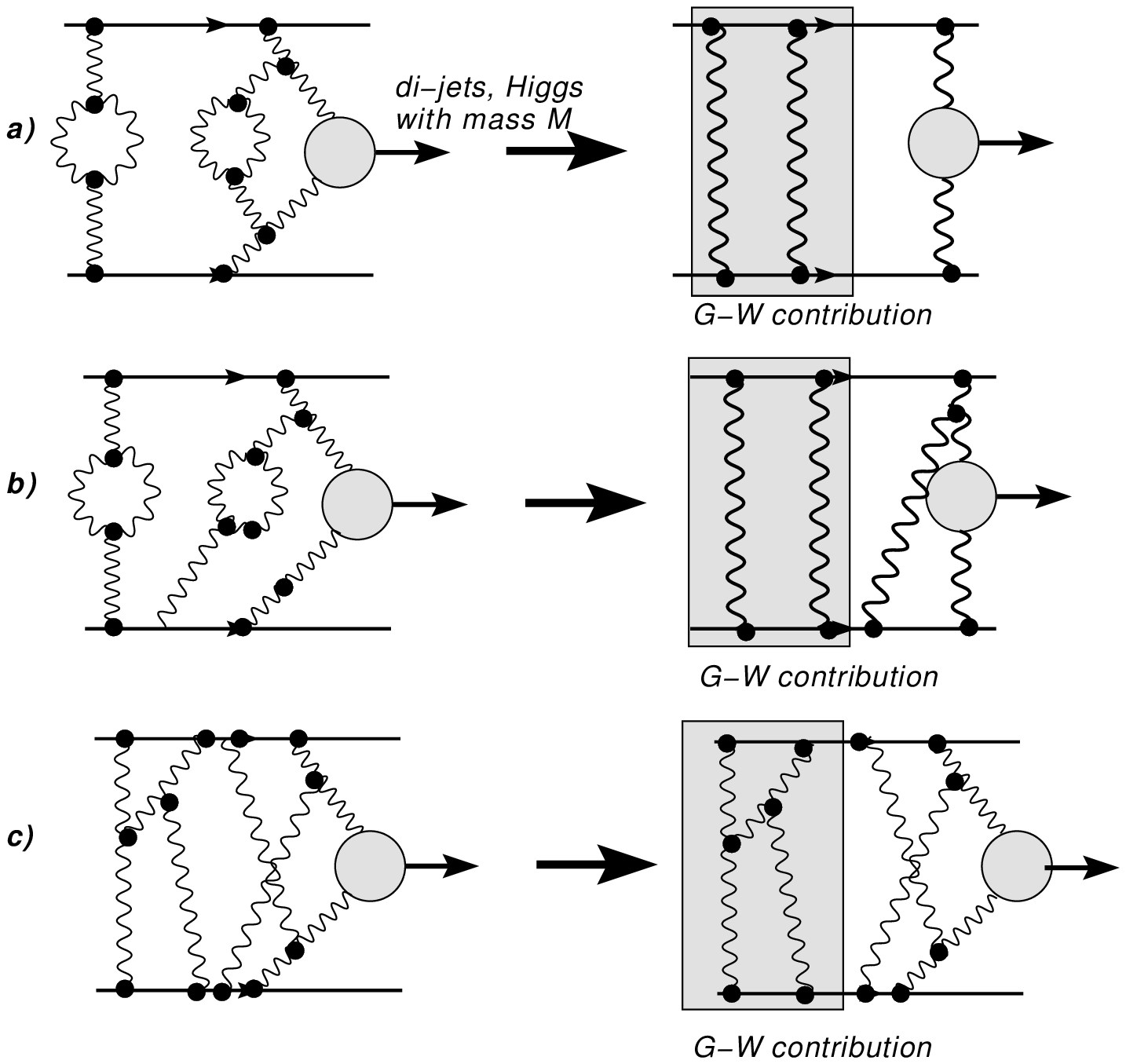,width=120mm}}
\caption{The set of diagrams that is selected and summed  for the calculation of the survival
 probability for diffractive Higgs production.\protect\fig{spn}-a shows the
 diagrams in G-W + enhanced diagrams approach, in \protect\fig{spn}-b  the same
 approach is shown, however, to calculate the value of
the survival probability
 we add the first semi-enhanced diagram.
  The approach for $\tilde{g}_i T(Y) \approx 1$ but $\Delta_\pom T(Y)
 \ll 1$ (net diagrams) is shown in \protect\fig{spn}-c.}
\label{spn}
\end{figure}

The factors $\exp[- \Omega_{i,k}/2]$ are responsible for the Good-Walker mechanism contribution for the 
survival 
probability, while the factors $S^H_{i,k}(b)$  account for an additional suppression
 due to a structure of the $\Omega_{i,k}$. In \fig{spn} we illustrate examples of
 the different contributions in $S^H_{i,k}$.
\FIGURE[ht]{
\centerline{\epsfig{file=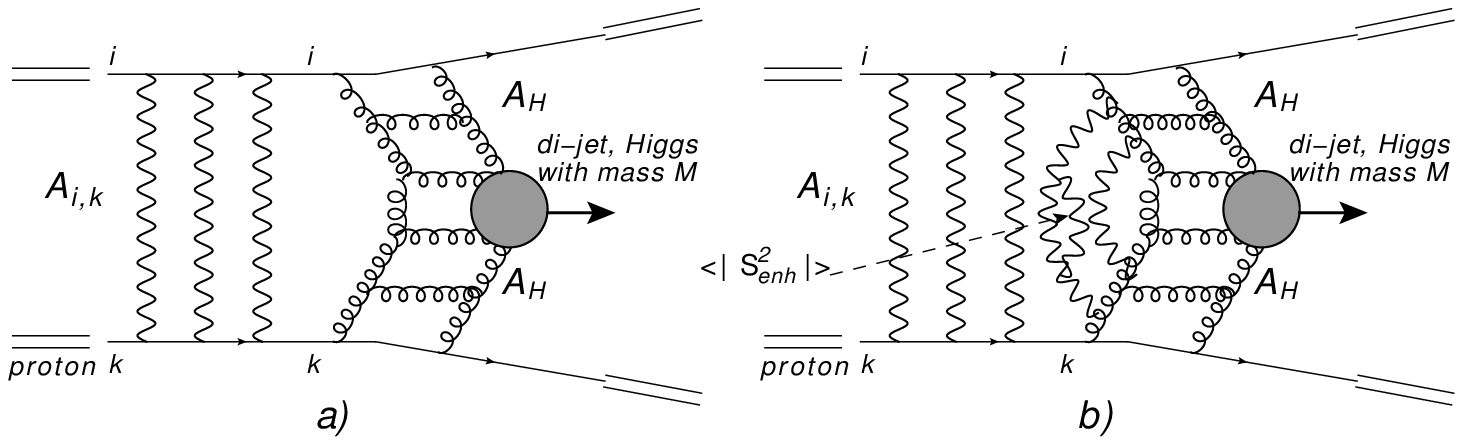,width=150mm}}
\caption{Survival probability for exclusive central diffractive
production of the Higgs boson. \fig{sp-dia}-a shows the contribution to
the survival probability in
the G-W mechanism, while \fig{sp-dia}-b  illustrates the origin of the
additional factor
$\langle\mid S^2_{enh}\mid\rangle$. }
\label{sp-dia}}
\fig{sp-dia}, which is taken from our paper \cite{GLMM} shows that
 $S^H_{i,k}$  depends critically on both the hard and the soft amplitude. Our main goal is
  to find this dependence and calculate  $S^H_{i,k}$.
 The parametrization of the  hard amplitude $A^i_H(s,b_i)$ has been discussed in Ref. 
\cite{GLMM}, 
 and we refer the reader to this paper for  details.

\section{Enhanced diagrams}

The calculation of the sum of the enhanced diagrams has been treated in detail in our previous
  papers\cite{GLMM,GLMLA} and so we will not repeat the discussion here.

In the case of summing the enhanced diagrams 
the damping factors $S^H_{i,k}(b;b_1b_2)$ have a simple structure
\beq \label{SPED1}
S^H_{i,k}(b;b_1b_2)\,\,\,=\,\,\,\langle\mid S_{enh}\Lb Y\Rb \mid \rangle\,\,A^i_H(s,b_1)\,A^k_H(s,b_2),
\eeq
where $\langle\mid S_{enh}\Lb Y\Rb \mid \rangle$ depends only on energy (total rapidity $Y=\ln(s/s_0)$) and 
it can be considered to be a common factor.
Although, this factor appears in Ref. \cite{GLMM} ,  
 for completeness of the  presentation it is  given below. 
\bea
&&\langle \mid S^2_{enh}\Lb Y\Rb \mid \rangle \,\,\,=  \label{SPE}\\
&&\,\,\,\,\,\,\,\,\,\,\,\,\,\,\,\,\,\,\,\,\,\,\,\,
= \,\,S\Lb {\cal T}( Y)\Rb \,\,-\,\,2\,
e^{ - \Delta_\pom (Y - y_h)/2}\,S1\Lb {\cal T}( Y)\Rb\,\,+\,\,
e^{ - 2\Delta_\pom (Y -  y_h)/2}\,S2\Lb {\cal T}( Y)\Rb; \notag\\
&& \,\,\,\,\,\,\,\,\,\,S(T)
= \frac{1}{T^3}\left\{ - T\,+\,\,\Lb 1  + T \Rb \,
e^{\frac{1}{T}}\,\Gamma\Lb 0,  \frac{1}{T}\Rb\right\}; \label{SPE51}\\
&& \,\,\,\,\,\,\,\,\,S1(T) =
\frac{1}{T^3}\left\{   -T (1 + T) \,+\, ( 1 + 2T)\,
e^{\frac{1}{T}}\,\Gamma\Lb 0, \frac{1}{T}\Rb\right\};\label{SPE52}\\
&& \,\,\,\,\,\,\,\,\,S2(T) =
\frac{1}{T^3} \left\{ T\left[ \Lb  T  - 1 \Rb^2 - 2 \right]  \,
+\, ( 1 + 3 T)\, e^{\frac{1}{T}}\,\Gamma\Lb 0,
\frac{1}{T}\Rb\right\},\label{SPE53}
\eea
where
\beq \label{TT}
{\cal T} \Lb Y\Rb\,\,\,=\,\,\gamma \Lb 
e^{\Delta_\pom ( Y - Y')} \,-\, 1\Rb\,\Lb e^{ \Delta_\pom Y'} \,-\, 1\Rb.
\eeq
 $\Gamma\Lb 0, 1/T\Rb$ is the incomplete gamma function
  (see formulae {\bf 8.35} in Ref.\cite{RY}).

\section{Enhanced diagrams and semi-enhanced diagrams as a  perturbation}
\begin{figure}[ht]
\centerline{\epsfig{file=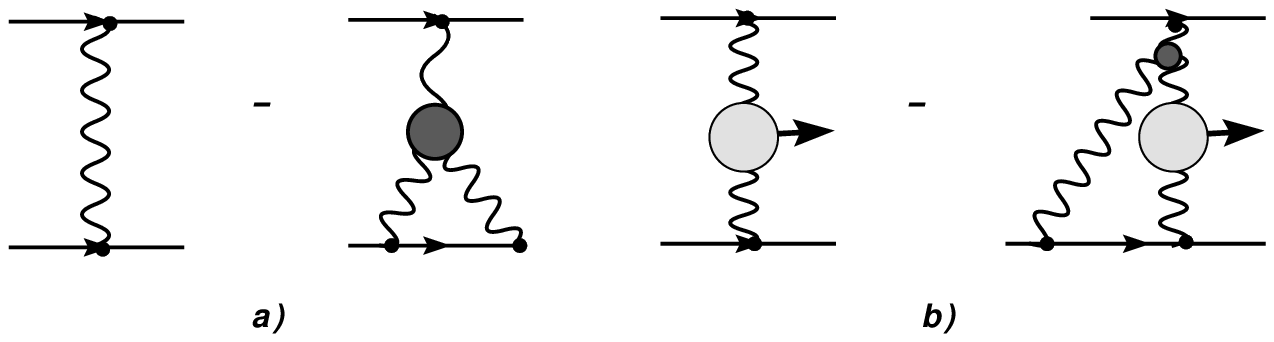,width=120mm}}
\caption{The set of diagrams  in terms of the exact Pomeron and exact vertices that we
 include in our perturbative approach for the summation of semi-enhanced diagrams.
 \protect\fig{seen}-a gives contribution to $\Omega_{i,k}$ (see section 2.3) while
 \protect\fig{seen}-b shows the changes in $S^H_{i,k}$.}
\label{seen}
\end{figure}
As has  been discussed in our approach based on the summation of the enhanced
 diagrams (see Ref. \cite{GLMM}) 
the contribution of the semi-enhanced diagrams to the value of the survival probability has been 
neglected.
 The first correction due to such a contribution is shown in \fig{spn}-b ( compare with \fig{spn}-a). 
 The expression for this correction 
has the form (see \fig{seen}-b)
\bea
&&\langle \mid S^H_{i,k} \Lb \fig{seen}-b;  Y,b \Rb \mid \rangle  \,\,\,\,=\,\,\,
 \int\,d^2 b_1 \left\{ \langle \mid S^2_{enh}\Lb Y\Rb \mid \rangle \, A^H_i\Lb Y,b_1\Rb \,
A^H_i\Lb Y,\vec{b} - \vec{b}_1\Rb\,\,\,\right. \label{EDSEP1}\\
&&\left.-\,\,\,2\,\,\int^{Y} d y_1\int^{y_1} d y_2 \int^{y_1} d y_3\,\,G\Lb Y - y_1\Rb\,
 \Gamma\Lb y_1,y_2,y_3\Rb\,G\Lb y_2 -0\Rb\,\langle \mid S^2_{enh}\Lb y_2\Rb \mid \rangle \,\right. \nonumber\\
&&\left. \times \,\,A^H_i\Lb Y,\vec{b} - \vec{b}_1\Rb\,\int\,d^2 b'
A^H_k\Lb Y,b'\Rb\,\tilde{g}_k\Lb \vec{b_1} - \vec{b'}\Rb\right\}.
\eea
The coefficient  2 stems from two diagrams where the exact Pomeron is attached to upper and lower proton  
in \fig{seen}-b. $G(Y)$ is the exact Green's function of the Pomeron and is given by
\beq \label{G}
G\Lb Y\Rb\,\,=\,\,1 \,-\,\\exp\Lb \frac{1}{T\Lb Y\Rb}\Rb\,\frac{1}{T\Lb Y\Rb}\,\Gamma\Lb 0,\frac{1}{T\Lb Y\Rb} \Rb
\eeq
with
\beq \label{ES11}
T\Lb Y \Rb\,\,\,=\,\,\gamma\,e^{\Delta_\pom Y}.
\eeq
 $ \Gamma\Lb y_1,y_2,y_3\Rb$ is the exact vertex that has been discussed in section 3.4 of Ref.\cite{GLMLA}.
It is equal to
\bea
&&\Gamma\Lb y1,y2,y3\Rb\,\,=\label{SD1} \\
&&= \frac{\Delta_\pom}{4} \,\frac{1}{T(y_1 - y_2) - T( y_1 - y_3)}\,\left\{ \Gamma_1\Lb 
2\,T\Lb y_1 - y_2\Rb\Rb\,\,-\,\,
 \Gamma_1\Lb 2\,T\Lb y_1 - y_3\Rb \Rb\right\}\\
&&\mbox{with}\,\,\, \Gamma_1\Lb T\Rb\,\,\,=\,\,\,(1/T^3)\times \left\{ T(1+T) - \exp\Lb - 1/T\Rb\,\Lb 1
 + 2 T\Rb \,\Gamma\Lb 0,1/T\Rb\right\}. \label{SD3}
\eea

 To calculate the survival probability
 we need to introduce the additional change in $\Omega_{i,k}$ in \eq{SP} 
  (see for example Ref. \cite{BORY} for detailed arguments).
 The resulting $ \Omega_{i,k}$
has the form (see \fig{seen}-a)
\bea
&& \Omega_{i,k}\Lb Y,b\Rb\,\,\,=\,\,\int\,d^2 b_1 \left\{ G\Lb Y\Rb \tilde{g}_{i}\Lb b_1\Rb\,\tilde{g}_{i}\Lb \vec{b} - \vec{b}_1\Rb \right. \label{EDSEP2}\\
&&\left.\,\,-\,\,2 \,\int^{Y} d y_1\int^{y_1} d y_2 \int^{y_1} d y_3\,\,G\Lb Y - y_1\Rb \,\Gamma \Lb y_1,y_2,y_3\Rb\,G\Lb y_2 -0\Rb\,\,G\Lb y_3 -0\Rb\,\right. \nonumber\\
&&\left. \times \,\,\tilde{g}_i\Lb \vec{b} - \vec{b}_1\Rb\,\int\,d^2 b'
\tilde{g}_k\Lb b'\Rb\,\tilde{g}_k\Lb \vec{b}_1 - \vec{b}'\Rb\right\}.
\eea

\section{Net diagrams.}
\begin{figure}[ht]
\centerline{\epsfig{file=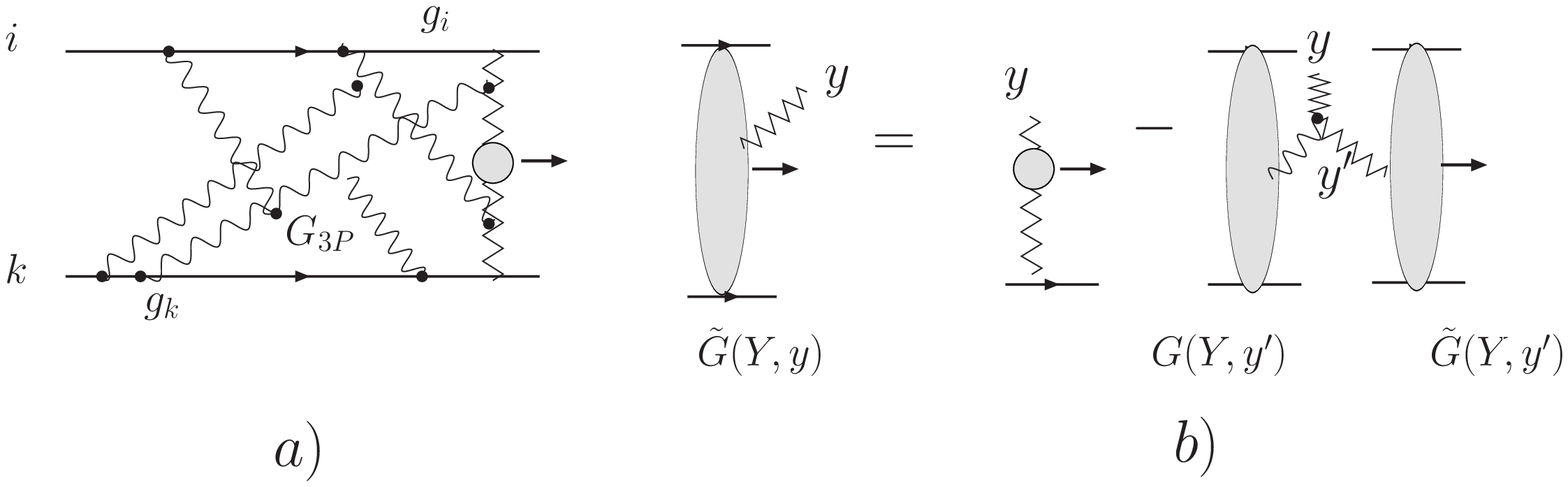,width=120mm}}
\caption{The set of diagrams( see \protect\fig{eqnd}-a)  for  $S^H_{i,k}$ in the approximation where $\tilde{g} Y\Lb Y\Rb \approx 1 $ while $\Delta T\Lb Y\Rb \ll 1$ ( net diagrams)  and the  graphic form of the equation for 
$S^H_{i,k} \equiv \tilde{G}_{i,k}$. Zigzag line denotes the hard Pomerons. }
\label{eqnd}
\end{figure}
  
In this section we  discuss the survival probability in the approximation
 where $\tilde{g}_i T\Lb Y\Rb \approx 1$ while $\Delta_\pom T\Lb Y\Rb \ll 1$. In this
 approximation we need to sum a `net' system of the diagrams, an example of which
 is shown in \fig{eqnd}-a. To find $S^H_{i,k}$ we introduce the function
$\tilde{G}_{k}\Lb Y;y \Rb$ for which we can write the equation (see \fig{eqnd}-b)
\beq \label{SPND1}
 \tilde{G}\Lb Y,y\Rb\,\,\,=\,\,\,A^H_k\Lb b_1\Rb\,\,-\,\,\Delta \int^y
 d y' \,G\Lb Y,y'\Rb\,\tilde{G}\Lb Y,y'\Rb.
\eeq
\eq{SPND1} can be rewritten in the differential form as
\beq \label{SPND2}
\frac{d  \tilde{G}\Lb Y,y\Rb}{ d y}\,\,\,= \,\,-\,\,\Delta\,\,G\Lb  Y,y'\Rb\,\tilde{G}\Lb Y,y'\Rb. 
\eeq
The solution to \eq{SPND2} is
\bea \label{SPND3}
 \tilde{G}\Lb Y,y\Rb\,\,\,&=&\,\,\,A^H_k\Lb b_1\Rb\,\,\exp\left\{- \int^y_0\, \Delta\,\,G\Lb  Y,y'\Rb\,d y'  \right\} \nonumber\\
  &=&\,\,\,\A^H_k\Lb b_1\Rb \frac{1}{1 \,+\,\tilde{g}_i(b_2)\,T\Lb Y\Rb \,\,+\,\,\tilde{g}_k\Lb b_1\Rb
 \,T\Lb y\Rb}. 
\eea

$S^H_{i,k}$ can be expressed with the aid of the  functions $\tilde{G}$ in the following way:

\bea \label{SPND4}
S^H\Lb Y,b\Rb\,\,\,&=&\,\,\,\int d^2 b_1 \,A^H_i(\vec{b} - \vec{b_1})\,A^H_k( b_1)\times \frac{1}{\Lb 1
 + \Lb G_{3\pom}/\gamma\Rb \Big( \tilde{g}_k(\vec{b} - \vec{b}_1)\,T\Lb Y\Rb \,\,+\,
\,\tilde{g}_i(b_1)\,T\Lb Y/2 - y_h/2\Rb \Big) \Rb}\nn\\
&\times & \frac{1}{ \,\Lb 1 + \Lb G_{3\pom}/\gamma\Rb \Big( \tilde{g}_i(\vec{b} -
 \vec{b}_1)\,T\Lb Y\Rb \,\,+\,\,\tilde{g}_k(b_1)\,T\Lb Y/2 +y_h/2\Rb\Big)\Rb},
\eea
where $y_h \,\,=\,\,\ln (M^2/s_0)$ and $M$ denotes the mass of the produced system (dijets,
 Higgs boson etc.) and  $s_0 = 1 \,GeV^2$.

Finally to calculate the survival probability we require
 $\Omega_{i,k}$ which was derived in Ref.\cite{GLMLA}

\beq \label{SPND5}
\Omega_{i,k} \,\,=\,\,\,\,\h\,  \int d^2 b'\,
\,\,\,\frac{\tilde{g}_i\Lb\vec{b}'\Rb\,\tilde{g}_k\Lb\vec{b} -
 \vec{b}'\Rb\,\Lb 1/\gamma\Rb T(Y)}{ 1\,+\,\Lb G_{3\pom}/\gamma\Rb T(Y)\,\left[\tilde{g}_i\Lb\vec{b}'\Rb
 + \tilde{g}_k\Lb\vec{b} - \vec{b}'\Rb\right]}.
\eeq

\section{Net and enhanced  diagrams.}
\begin{figure}[ht]
\centerline{\epsfig{file=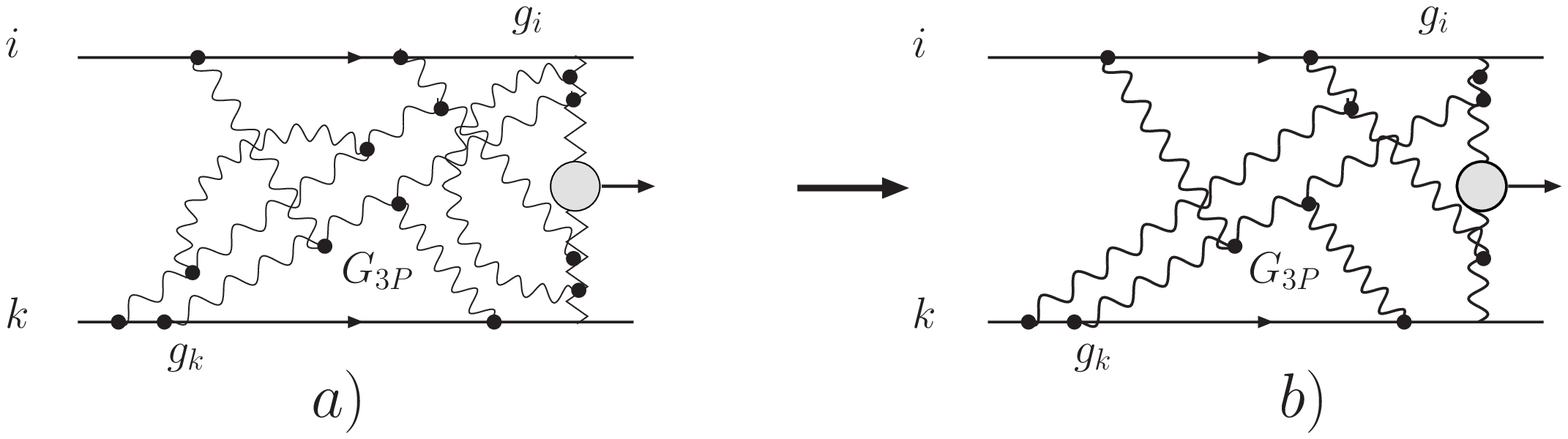,width=140mm}}
\caption{The net and enhanced diagrams (\protect\fig{eqnden}-a)  and the set of diagrams
 with the exact Green's function of  Pomeron and exact vertices (\protect\fig{eqnden}-b).
 Zigzag line denotes the hard amplitudes. }
\label{eqnden}
\end{figure}
In the spirit of our approach we can generalized \eq{SPND4} and \eq{SPND5} taking into account the 
enhanced diagrams. As we have discussed, we need to replace $T\Lb Y \Rb $ by $G\Lb T\Lb Y\Rb \Rb$ 
in these equations (see \fig{eqnden}-a and \fig{eqnden}-b) and to
 multiply \eq{SPND4} by $\langle \mid S^2_{enh}\Lb Y\Rb \mid \rangle$. 
  This factor is responsible for summing the diagrams where the Pomeron lines are
 attached to the hard amplitude (`hard Pomeron') (see \fig{eqnden}-a). 

\section{Numerical estimates and prediction for the LHC}

\begin{figure}[ht]
\centerline{\epsfig{file=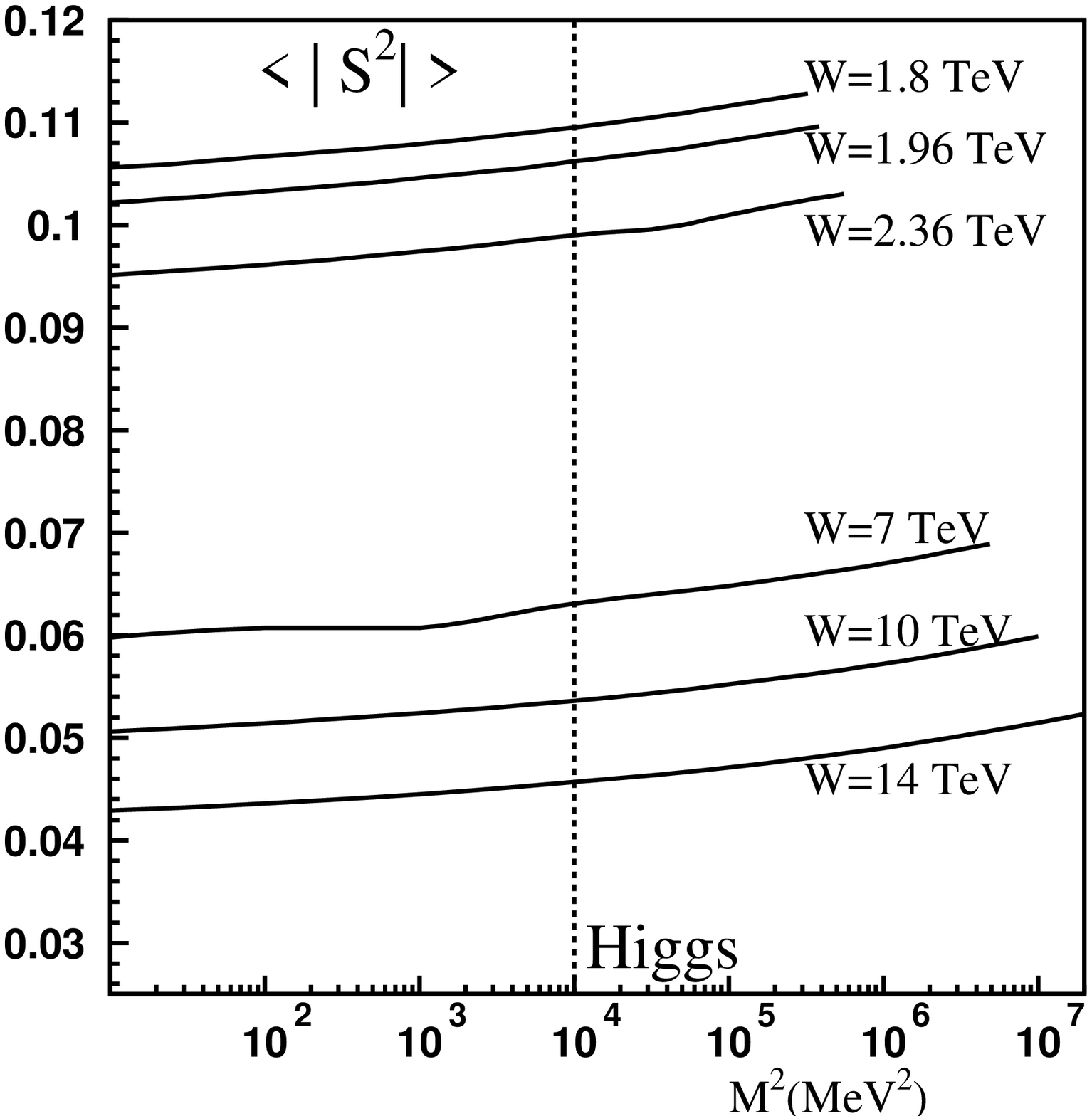,width=90mm}}
\caption{The estimates for the value of SP  for  dijets with mass $M$. The dotted line corresponds
 to the Higgs boson production with $M_{higgs} = 100 \,GeV$. We estimate a  $60 \div 70 \% $      margin of error in our results. }
\label{sp}
\end{figure}

Using \eq{SPND4} and \eq{SPND5} we have calculated the survival probability as a function of produced di-jets mass at c.m. energies of interest. We estimate the margin of errors of our results to range from 60\%  at W=2 TeV to 70\% at W=14 TeV. The central values obtained are shown in \fig{sp}.

The main feature of our SP results is that their values are considerably larger than the values obtained in our previous studies\cite{GLMLA}. Our current estimates are compatible with the CDF Tevatron data \cite{CDF} as well as with the theoretical estimates \cite{KMRCDF, ROYON}. 

Our calculations presented in Ref.\cite{GLMM} only summed over the enhanced diagrams, while our present calculations sum over the complete set of enhanced, semi-enhanced and net diagrams. Comparing the fitted values of our parameters in the two studies we note that our present fitted values of $\gamma$ and $G_{3 \pom}$ are much smaller than those obtained in Ref. \cite{GLMM}. Consequently, the screening initiated by the enhanced diagrams with the present parameters is much smaller that the screening caused by the same diagrams in our previous study. 
On the other hand, in our both papers we found out that (i) the main source of the small values of the  SP
is the enhanced diagrams; and (ii) the enhanced diagrams are the only cause for the dependence of the SP value on the mass of produced di-jets. Therefore,  the rather large value of  SP and its  mild dependence of the mass of produced di-jets have the same origin: the small values of $G_{3\pom}$.

\section{Conclusions}
\fig{sp} shows our estimates and predictions for the value of the SP. These values are the  results 
of our 
theoretically self consistent  approach that has been discussed in Refs.\cite{GLMM,GLMLA}. 
  We predict the value of the SP for Higgs production ($M_{Higgs}=100\,GeV$) of about $3-5\%$ for
 the LHC range of energies. These values are much larger our previous estimates, for a model 
in which
 only enhanced diagrams were taken into account \cite{GLMM}.
  Our main conclusions are  that
 the value of the SP, as well as its mass dependence are very
 sensitive to both  the particular form adopted for the Pomeron interaction, and 
to the values of the fitted parameters that determine the strength of the   contribution of the
 enhanced diagrams.

The lesson from our present  study is straight forward and discouraging
to an extent, since we need a thorough knowledge of  the theory of interacting
 Pomerons  to guarantee the value of the SP.  On the other hand, the
    model we constructed
incorporates  the main features of two compelling  theoretical approaches:
 N=4 SYM and perturbative QCD.

We conclude that  our theoretical knowledge of Pomeron 
interactions, as well as the set of available experimental data,
 are not sufficient to determine the strength
 of the Pomeron interaction with adequate
  accuracy,  so as to provide   precise estimates for the SP.


\end{document}